\documentclass[aps,pra,reprint,twocolumn,floatfix,groupedaddress,superscriptaddress]{revtex4-2}
\usepackage{pstricks,graphicx,amsmath,bbm,mathrsfs,amssymb,psfrag,pifont,times,mathptmx}
\usepackage[utf8]{inputenc}
\usepackage{hyperref}
\usepackage[english]{babel}
\usepackage{color}
\usepackage{ulem}
\usepackage{siunitx}
\usepackage{mathtools}
\usepackage{comment} 
\newcommand{\NOTll}{\hskip 0.4mm \not \hskip -0.4mm \ll}
\def\Id{{\mathbbm 1}}
\DeclareMathOperator{\Tr}{Tr}
\def\dag{{^{\dagger}}}
\def\Re{{\mathrm{Re}}}
\def\Im{{\mathrm{Im}}}
\def\id{ {\mathrm{id}} }
\def\a{ {\mathrm{QS}} }
\def\b{ {\mathrm{SPC}} }
\def\p{ {\mathrm{p}} }
\def\sigmamA{\boldsymbol\sigma^{\rm(m)}_{A}}
\def\sigmamB{\boldsymbol\sigma^{\rm(m)}_{B}}
\def\sigmamAB{\boldsymbol\sigma^{\rm(m)}_{A(B)}}
\def\sigmaz{\boldsymbol\sigma_z}
\def\aA{a_A}
\def\aB{a_B}
\def\aC{a_{B_1}}
\def\aD{a_{B_2}}
\def\bA{b_A}
\def\bB{b_B}
\def\bC{b_{B_1}}
\def\bD{b_{B_2}}
\def\alphaA{\alpha_A}
\def\alphaB{\alpha_B}
\def\alphaC{\alpha_{B_1}}
\def\alphaD{\alpha_{B_2}}
\def\betaA{\beta_A}
\def\betaB{\beta_B}
\def\betaC{\beta_{B_1}}
\def\betaD{\beta_{B_2}}\def\alphaA{\alpha_A}
\def\alphaB{\alpha_B}
\def\alphaC{\alpha_{B_1}}
\def\alphaD{\alpha_{B_2}}
\def\betaA{\beta_A}
\def\betaB{\beta_B}
\def\betaC{\beta_{B_1}}
\def\betaD{\beta_{B_2}}
\def\bmalpha{\boldsymbol\alpha}
\def\bmbeta{\boldsymbol\beta}
\begin{document}
\title{Long-distance continuous-variable quantum key distribution with feasible physical noiseless linear amplifiers}
\author{Michele N.~Notarnicola}
\affiliation{Dipartimento di Fisica ``Aldo Pontremoli'',
Universit\`a degli Studi di Milano, I-20133 Milano, Italy}
\affiliation{INFN, Sezione di Milano, I-20133 Milano, Italy}
\author{Stefano Olivares}
\email{stefano.olivares@fisica.unimi.it}
\affiliation{Dipartimento di Fisica ``Aldo Pontremoli'',
Universit\`a degli Studi di Milano, I-20133 Milano, Italy}
\affiliation{INFN, Sezione di Milano, I-20133 Milano, Italy}
\date{\today}
\begin{abstract}
Noiseless linear amplifiers (NLAs) provide a powerful tool to achieve long-distance continuous-variable quantum key distribution (CV-QKD) in the presence of realistic setups with non unit reconciliation efficiency. We address a NLA-assisted CV-QKD protocol implemented via realistic physical NLAs, namely, quantum scissors (QS) and single-photon catalysis (SPC), and compare their performance with respect to the ideal NLA $g^{\hat{n}}$. We investigate also the robustness of two schemes against inefficient conditional detection, and discuss the two alternative scenarios in which the gain associated with the NLA is either fixed or optimized.
\end{abstract}
\maketitle

\section{Introduction}
Quantum key distribution (QKD) \cite{RevModPhys.74.145} allows to share a common secure key between a sender and a receiver even in the presence of an untrusted channel that could be under the control of an eavesdropper. Within this framework, a promising role is played by continuous-variable QKD (CV-QKD) for both theoretical and experimental reasons \cite{RevModPhys.77.513}. In the first proposal of a CV-QKD scheme by Grosshans and Grangier (GG02) \cite{Grosshans_Grangier_2002, Grosshans_2003, Weedbrook_2004, Grosshans_Cerf_Wenger2003, Grosshans_2005} information is encoded by the sender (Alice) on the quadratures of a quantized optical field with Gaussian modulation and then sent into a channel to the receiver (Bob) that performs either homodyne or heterodyne (double-homodyne) measurements. The key is then extracted after a reconciliation process, where one of the two parties publicly reveals part of the data: if such party is Alice the process is referred to as direct reconciliation, if the party is Bob we have reverse reconciliation.
The security analysis of the reverse-reconciliation protocol guarantees a non null secure key rate for any transmission distance \cite{Grosshans_Grangier_2002, Grosshans_2003,Grosshans_2005,PirandolaRev}. 
\par
In realistic conditions, however, the reconciliation procedure is not perfect and one can introduce a reconciliation efficiency, which depends on the particular code employed to extract the secure key \cite{Bloch_Thangaraj_McLaughlin_Merolla_2006}. Moreover, the presence of defects inside Alice's Gaussian modulator as well as phase noise of the carrier signal introduce an excess noise \cite{Lodewyck_2005}. Both these limitations crucially affect the key generation rate (KGR), i.e. the length of the secret key shared by Alice and Bob per unit time slot, and prevent long-distance communication leading to a maximum transmission distance at which the KGR vanishes \cite{Leverrier_2008,Lodewyck_2005, NotarnicolaShap}. 
In the latest experimental realizations, high-loss CV-QKD has been achieved up to maximum transmission distances ranging from $100$ to $200$~km \cite{Jouguet2013, Zhang2020, Pi2023, Bian2023, Hajomer2023}.
\par
A challenging task to face those issues is to modify the original protocol by implementing strategies allowing to increase as much as possible the maximum transmission distance. An intriguing solution is provided by heralded noiseless linear amplification at the receiver's side \cite{NLARalphLund, RalphREP, Adnane2019}. Indeed, an ideal probabilistic noiseless linear amplifier (NLA) with amplitude gain $g$ leads to an increase in the maximum transmission distance proportional to $\log g$ \cite{Blandino}. Nevertheless, any realistic physical NLA can only approximate the ideal amplifier for low-amplitude optical signals \cite{NLARalphLund, NLAFiu1, NLAXi, NLAMc, NLASPC, NLARalph, NLAFiu2, NLAJoshua1, NLAFiu3, NLAJoshua2}. To avoid this limitation, measurement-based NLAs, performing virtual amplification based on classical data post-selection, have also been proposed \cite{MB1, MB2, MB3}. However, the low success probabilities of these operations  \cite{Bernu_2014, Zhao_2017} make physical NLAs still worth of investigation. Recently, CV-QKD employing quantum scissors (QS) \cite{NLARalphLund} has been addressed, allowing to achieve long-distance CV-QKD for sufficiently low channel excess noise \cite{Ghalaii,Ghalaii_DM}. To the same goal, also single-photon catalysis (SPC) has been investigated \cite{NLASPC, QKDNonGauss}.
In the QS scheme, a single photon is mixed with the vacuum at a beam splitter with transmissivity $\tau$. One of the output branches then impinges at a balanced beam splitter with the incoming signal, after which double conditional photo-detection is performed. Differently from QS, in the SPC process a single photon interferes directly with the incoming signal at a beam splitter with transmissivity $\tau$ and then a single photon is retrieved at the end. Thus, SPC provides a simpler scheme and may represent a feasible alternative to QS for experimental realizations. 
\par
In the following paper we investigate a CV-QKD protocol assisted by these two schemes and consider a simplified realistic scenario, where photo-detection is replaced by on-off detection. We compute the KGRs for both the strategies and compare them to the performance of the protocol assisted by the ideal NLA proposed in \cite{Blandino}. Moreover, we distinguish two alternative cases. In the former, we fix the NLA gain $g$ and show that also physical NLAs increase the maximum transmission distance by the same amount $\log g$ as the ideal amplifier. In the latter, we assume $g$ to be a free parameter and optimize its value, obtaining that both physical and ideal NLAs achieve arbitrary long-distance CV-QKD.
For the physical amplifiers, we also discuss the robustness in the presence of a quantum detection efficiency $\eta\le1$, showing that the detection efficiency only rescales the KGR without preventing long-distance communication.
\par
The structure of the paper is the following. In Sec.~\ref{sec: GG02} we recall the main features of the GG02 protocol. Then, in Sec.~\ref{sec: NLA} we describe the NLA-assisted protocols for both the ideal and the physical amplifiers, namely, QS and SPC. In Sec.~\ref{sec: Security} we perform the security analysis by comparing the KGRs of the protocols under investigation. Finally, in Sec.~\ref{sec: Concl} we summarize the results obtained and draw some conclusions.

\section{The GG02 original protocol}\label{sec: GG02}
We start reviewing the CV-QKD protocol proposed in \cite{Grosshans_Grangier_2002, Grosshans_2003, Weedbrook_2004,Grosshans_Cerf_Wenger2003} in its entanglement-based (EB) version, which provides a simplified theoretical analysis  \cite{GarciaPatronCerf_2009, Lodewyck_2007}.
Here, Alice and Bob share a two-mode squeezed vacuum state (TMSV), namely,
$
|{\rm TMSV} \rangle\!\rangle =
\sqrt{1-\lambda^2}\sum_{n=0}^{\infty} \lambda^n |n\rangle |n \rangle
$
with $0\le \lambda \le 1$.
The TMSV is a two-mode Gaussian state \cite{Olivares_PLA, Olivares_2012}, and can be completely described by the covariance matrix (CM) (see Appendices~\ref{app: Gauss} and \ref{app: EBGG02} for details)
\begin{align}\label{eq: Gamma EPR}
    \Gamma_{\rm TMSV} = \begin{pmatrix} V \, \Id_2 & Z \, \sigmaz \\ Z \, \sigmaz & V \, \Id_2 \end{pmatrix} \, ,
\end{align}
where $V = 1+ 2\lambda^2/(1-\lambda^2)$ is the TMSV variance, corresponding to the input modulation variance of the protocol, $Z=\sqrt{V^2-1}$, $\Id_2 = {\rm Diag}(1,1)$ and $\sigmaz$ is the Pauli $z$-matrix. All quantities are expressed in shot noise units.
\par
Now, Alice performs a heterodyne (i.e. double-homodyne) measurement on her beam, while the other one is sent to Bob through an untrusted communication channel, described by means of a thermal-loss channel.
The channel has a transmissivity $T= 10^{-\kappa d/10}$, where $d$ is the transmission distance in $\rm km$ and $\kappa \sim 0.2 \, \rm{dB/km}$ is the typical loss parameter for optical fibers at $\SI{1550}{\nano\meter}$ \cite{1550_1,1550_2,1550_3}. 
Moreover, a single-mode thermal bath of $n_{\epsilon}= T \epsilon/2(1-T)$ photons models the presence of an excess noise $\epsilon$ introduced by the realistic defects of Alice's modulation system \cite{Lodewyck_2005}.
Losses and imperfections affect the signal received by Bob that exhibits an added noise $\chi= (1-T)/T + \epsilon$, leading to an overall thermal-loss channel.
Therefore, the state shared between Alice and Bob is still Gaussian with CM \cite{Olivares_PLA, Olivares_2012}:
\begin{align}\label{eq: Gamma_GG02}
    \Gamma_{AB} = \begin{pmatrix} \Gamma_A & \Gamma_Z  \\ \Gamma_Z^\mathsf{T} & \Gamma_B \end{pmatrix} = \begin{pmatrix} V \, \Id_2 & \sqrt{T} Z \, \sigmaz \\ \sqrt{T} Z \, \sigmaz & T(V + \chi) \, \Id_2 \end{pmatrix} \, .
\end{align}
Once received the signal, Bob implements a Gaussian measurement \cite{GarciaPatronCerf_2009, Lodewyck_2007} that here we assume to be homodyne detection of a quadrature randomly chosen between $q$ and $p$, as in the original proposal \cite{Grosshans_Grangier_2002, Grosshans_2003}.
\par
All the necessary information to perform the security analysis is contained in the CM (\ref{eq: Gamma_GG02}). According to the Gaussian formalism \cite{Ferraro_Olivares_Paris_2005,Olivares_2012} when Alice and Bob perform detection on their own signals they get a bi-variate Gaussian distribution $p_{A(B)}(x_{A(B)}, y_{A(B)})$ with zero mean and covariance $\Gamma_{A(B)}+ \sigmamAB$, where $\sigmamA= \Id_2$ is the CM of the heterodyne detection and
\begin{equation}
\sigmamB = \lim_{z\rightarrow 0} \begin{pmatrix} z & 0 \\ 0 & z^{-1}\end{pmatrix}
\end{equation}
is the $2 \times$ CM associated with homodyne detection still in shot noise units (see Appendix~\ref{app: EBGG02}). 
Therefore, the joint measurement leads to the distribution $p_{AB}(x_A,y_A;x_B,y_B)$ with covariance $\Gamma_{AB}+(\sigmamA  \oplus \sigmamB )$. The mutual information between Alice and Bob is then given by:
\begin{align}\label{eq: IAB GG02}
    I_{AB} &= H[p_A]+H[p_B] - H[p_{AB}] \notag \\
    &= \log_2 \Bigg\{ \sqrt{\frac{\det\big[\Gamma_A+\sigmamA\big] \det\big[\Gamma_B+\sigmamB \big]}{\det\big[\Gamma_{AB}+(\sigmamA  \oplus \sigmamB )\big]}} \Bigg\} \, ,
\end{align}
$H[p]=-\int dx\, p(x) \log_2p(x)$ being the Shannon entropy of $p(x)$.
\par
Throughout this paper we will focus on a reverse reconciliation scheme, which has been proved to guarantee higher security than direct reconciliation \cite{Grosshans_2005, PirandolaRev}. Furthermore, we will assume an eavesdropper (Eve) to be able to perform collective attacks, which represent the best possible kind of attacks in his power, at least in the asymptotic limit of an infinite dataset \cite{Grosshans_2005}. If the reconciliation efficiency is $0\le \beta\le 1$, the KGRwrites
\begin{align}\label{eq: KGR GG02}
    K= \beta I_{AB}- \chi_{BE} \, ,
\end{align}
where the Holevo information $\chi_{BE}$ represents the amount of information extracted by Eve  \cite{Holevo} and can be computed starting from the CM~(\ref{eq: Gamma_GG02}) as:
\begin{align}\label{eq: chiBE GG02}
    \chi_{BE}= G\bigg(\frac{d_1-1}{2}\bigg)+G\bigg(\frac{d_2-1}{2}\bigg)- G\bigg(\frac{d_3-1}{2}\bigg) \, ,
\end{align}
where 
\begin{align}
G(x)= (x+1) \log_2 (x+1) - x \log_2 x \, ,
\end{align}
and $d_{1(2)}$ are the symplectic eigenvalues of $\Gamma_{AB}$ \cite{Ferraro_Olivares_Paris_2005, Olivares_2012}, namely
\begin{align}
    d_{1(2)}= \sqrt{\frac{\Delta \pm \sqrt{\Delta^2-4 I_4}}{2}} \, ,
\end{align}
with $I_{1(2)}= \det(\Gamma_{A(B)})$, $I_3= \det(\Gamma_Z)$, $I_4= \det(\Gamma_{AB})$ and $\Delta= I_1+I_2+2I_3$. Finally, $d_3 = \sqrt{\det(\Gamma_{A|B})}$ with (see Appendix~\ref{app: EBGG02}):
\begin{align}
    \Gamma_{A|B} = \Gamma_A - \Gamma_Z \bigg[ \Gamma_B + \sigmamB \bigg]^{-1} \Gamma_Z^\mathsf{T} \, .
\end{align}
\par
In the following we will study the behavior of $K$ as a function of the transmission distance $d$, optimizing over the modulation variance $V$ for fixed reconciliation efficiency $\beta \sim 0.95$ \cite{ Bloch_Thangaraj_McLaughlin_Merolla_2006, Leverrier_2009, Joguet} and the channel excess noise $\epsilon$.
\par
For the sake of clarity, we will review the results for the original protocol in the next section together with the NLA-assisted strategies under investigation.

\section{NLA-assisted CV-QKD}\label{sec: NLA}
In this section we investigate the performance of the CV-QKD protocol presented in Sec.~\ref{sec: GG02} assisted by a NLA. That is, Alice prepares the TMSV state with modulation variance $V$ and injects one mode into the thermal-loss channel. To mitigate the added noise $\chi$, Bob implements a NLA on his received pulse, before performing homodyne detection. 
\par
Here we consider Bob to employ either the ideal NLA proposed in \cite{Blandino}, or feasible physical NLAs realized via QS or SPC.

\subsection{Ideal NLA}
\begin{figure}[t]
\includegraphics[width=0.99\columnwidth]{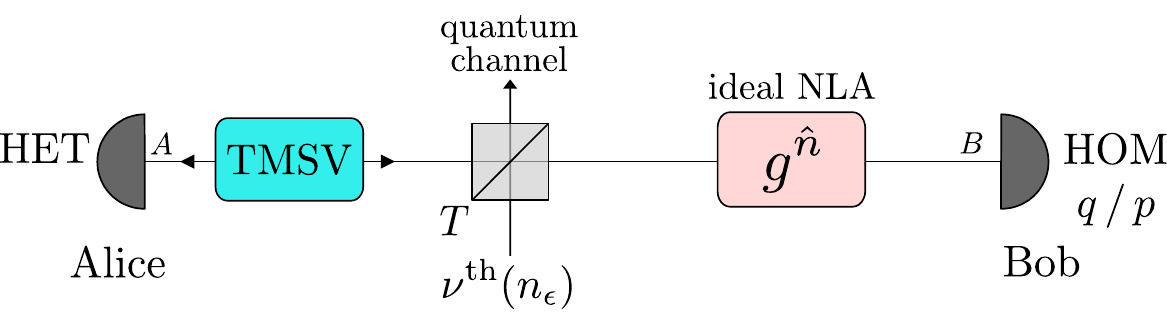}
\caption{Scheme of the CV-QKD protocol assisted by the ideal NLA proposed in \cite{Blandino}.}\label{fig:01-IdealNLA}
\end{figure}
At first, we assume Bob to employ an ideal NLA, as depicted in Fig.~\ref{fig:01-IdealNLA}. 
The ideal NLA is a non-deterministic operation described by the self-adjoint operator $g^{\hat{n}}$, $\hat{n}$ being the photon-number operator of the optical mode undergoing amplification, and $g \ge 1$ is the amplifier gain \cite{NLARalphLund}.
As discussed in \cite{Blandino}, this operation preserves Gaussianity, therefore the protocol in Fig.~\ref{fig:01-IdealNLA} is equivalent to a GG02 scheme with the following parameters:
\begin{subequations}\label{eq: IdealPar}
\begin{align}
V_{\id}&= V+ \frac{T (g^2-1) Z^2}{2-T (g^2-1)(V-1+\epsilon)} \, ,\\[1ex]
T_{\id} &=\frac{g^2 T}{1+ T (g^2-1)[1+T \epsilon (g^2-1)(2-\epsilon)/4-\epsilon]} \, ,\\[1ex]
\epsilon_{\id} &= \epsilon + (g^2-1) \frac{T \epsilon(2-\epsilon)}{2}  \,,
\end{align}
\end{subequations}
provided that:
\begin{align}\label{eq: Condition}
g \le \sqrt{1+\frac{2}{T(V+\epsilon-1)} } \,.
\end{align}
Without the last condition on the gain an unphysical un-normalizable state is obtained \cite{Blandino,NLARalphLund}. Equivalently, for a fixed gain Eq.~(\ref{eq: Condition}) corresponds to a threshold of the transmissivity, namely:
\begin{align}
T \le T_{\rm th} \equiv \frac{2}{(g^2-1)(V+\epsilon-1)} \, ,
\end{align}
preventing the use of the NLA protocol for distances $d \le~d_{\rm th}^{(\id)} = (-10 \log_{10}T_{\rm th})/\kappa$. For $d > d_{\rm th}^{(\id)}$, employing the ideal NLA is equivalent to considering an effective channel of increased transmissivity $T_{\id} \ge T$.
The resulting KGR then reads:
\begin{align}
    \widetilde{K}_{\id}(V,g)= P_{\id}(V,g) \bigg[\beta I_{AB}^{(\id)}(V,g)- \chi_{BE}^{(\id)}(V,g) \bigg] \, ,
\end{align}
where $P_{\id}(V,g)$ is the success probability of the NLA, whereas $I_{AB}^{(\id)}(V,g)$ and $\chi_{BE}^{(\id)}(V,g)$ are computed from Eq.s~(\ref{eq: IAB GG02}) and~(\ref{eq: chiBE GG02}), respectively, with the modified parameters~(\ref{eq: IdealPar}). Since $P_{\id}(V,g)\le 1/g^2$ \cite{Blandino}, from now on we consider as a benchmark the KGR
\begin{align}\label{eq: KGR ideal}
    K_{\id}(V,g)= \frac{1}{g^2} \bigg[\beta I_{AB}^{(\id)}(V,g)- \chi_{BE}^{(\id)}(V,g) \bigg] \, .
\end{align}
\par
The KGR~(\ref{eq: KGR ideal}) depends on the two free parameters $V$ and $g$ that can be optimized. As discussed in the rest of the paper, the choice of the gain $g$ will be a crucial task. Hence, we will discuss two separate cases. In the former case we assume a fixed $g$ and optimize only the modulation variance, obtaining the KGR
\begin{align}\label{eq: KGR ideal Opt1}
    K_{\id}(g)= \max_V K_{\id}(V,g) \, ,
\end{align}
and the corresponding distance-dependent modulation $V_{\rm opt}^{(\id)}(g)$. In the latter case the optimization involves also the gain, obtaining
\begin{align}\label{eq: KGR ideal Opt2}
    K_{\id}= \max_{V,g} K_{\id}(V,g) \, ,
\end{align}
and the associated parameters $V_{\rm opt}^{(\id)}$ and $g_{\rm opt}^{(\id)}$.

\subsection{Physical NLAs: QS and SPC}
\begin{figure}[t]
\includegraphics[width=0.99\columnwidth]{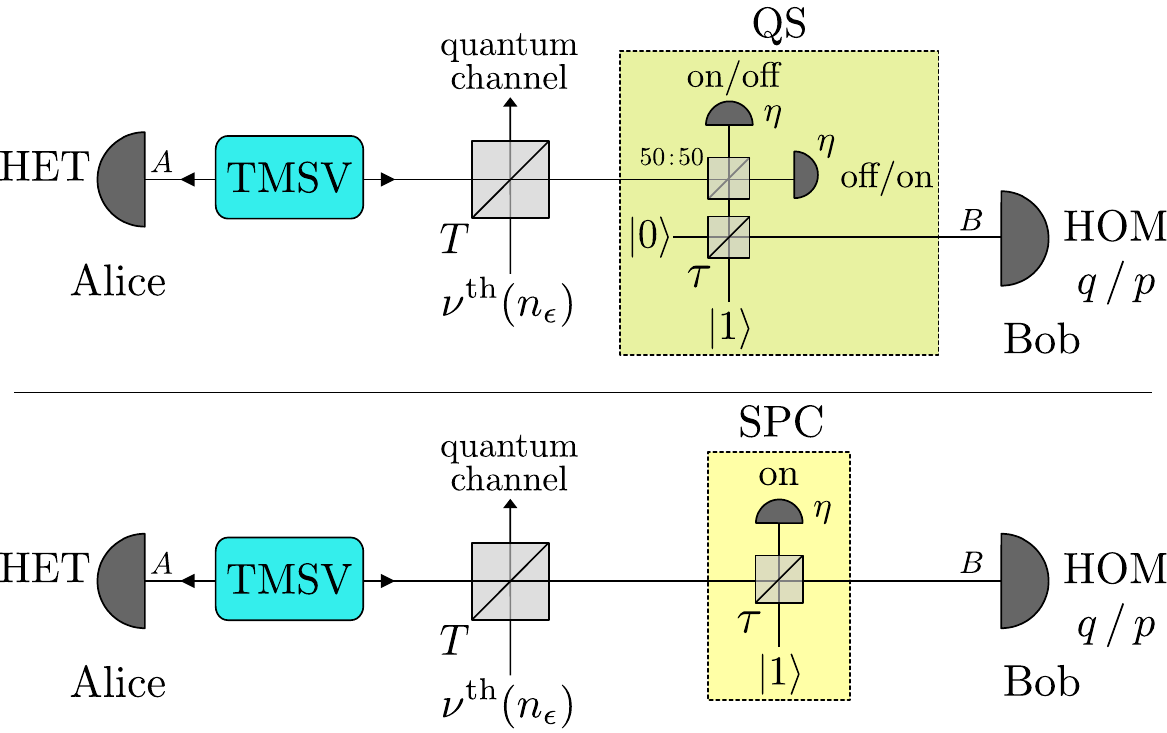}
\caption{Scheme of the CV-QKD protocol assisted by the two physical NLAs discussed in the paper. (Top) Strategy based on quantum scissors (QS); (bottom) strategy based on single-photon catalysis (SPC).}\label{fig:02-PhysicalNLA}
\end{figure}
Here we consider the more realistic scenario in which Bob employs a physical NLA, realized via either QS or SPC and employing on-off detection rather than photon counting.
\par
In the QS scheme proposed in \cite{Ghalaii} (Fig.~\ref{fig:02-PhysicalNLA}, top panel), Bob prepares two ancillary modes in the Fock states $|1\rangle$ and $|0\rangle$, respectively. He mixes them at a beam splitter with transmissivity $\tau$ and lets the reflected signal interfere at a balanced beam splitter with the pulse received by Alice. Then, he performs conditional on-off detection on both the output branches (see Appendix~\ref{app: NLA} for details), corresponding to the positive-operator-valued measurement (POVM) $\{\Pi_{\rm off}, \Pi_{\rm on}=\Id-\Pi_{\rm off}\}$, where 
\begin{equation}
\Pi_{\rm off} = \sum_{k=0}^{\infty} (1-\eta)^{k} |k\rangle \langle k | \, ,
\end{equation}
and $\eta\leq 1$ is the detection quantum efficiency. If one of the two detectors gives the outcome ``on", Bob performs homodyne detection on the post-selected output state. The value of $\tau$ fixes the gain associated with the NLA, that for low-amplitude coherent signals reads $g= \sqrt{(1-\tau)/\tau}$ \cite{NLARalphLund}. 
Thus, to achieve the gain $g$ we set the transmissivity equal to
\begin{align}\label{eq: gQS}
\tau_{\a}(g)= \frac{1}{1+g^2} \, .
\end{align}
\par
On the contrary, in the SPC scheme (Fig.~\ref{fig:02-PhysicalNLA}, bottom panel), Bob has a single ancillary mode excited in $|1\rangle$ impinging at a beam splitter with transmissivity $\tau$ with the pulse received by Alice. He performs on-off detection on the reflected branch, conditioning on outcome ``on", and homodynes the post-selected state. The associated gain is $g= (1-2\tau)/\sqrt{\tau}$ \cite{NLASPC}, which can be inverted to find the transmissivity as a function of the gain
\begin{align}\label{eq: gSPC}
\tau_{\b}(g)= \frac18 \bigg( 4 + g^2 - g \sqrt{8 + g^2}\bigg) \, .
\end{align}
\par
In both the cases, after the NLA Alice and Bob share a non-Gaussian state $\rho_{AB}^{(\p)}$, $\p= \a,\b$. 
However, since Bob's measurement is Gaussian, the security analysis of the NLA-assisted protocol can be based on the optimality of Gaussian attacks \cite{GaussOpt1, GaussOpt2, GaussOpt3}, which, in this scenario, maximize the amount of information extractable by Eve.
Moreover, following Ref.~\cite{GaussOpt1}, we consider the Gaussian lower bound on the mutual information, that is a consequence of the Gaussian (heterodyne) detection at Alice's side. In turn, we can compute a \textit{lower bound} of the exact KGR as:
\begin{align}\label{eq:K NotOpt}
K_{\p}(V,g) = P_{\p}(V,g) \bigg[ \beta I_{AB}^{(\p)}(V,g) -\chi_{BE}^{(\p)}(V,g) \bigg] \, ,
\end{align}
where $P_{\p}(V,g)$ is the success probability associated with the $\p$-th NLA and $I_{AB}^{(\p)}(V,g) $ and $\chi_{BE}^{(\p)}(V,g)$ are the mutual information and the Holevo information, respectively, both computed for a Gaussian state having the same CM of $\rho_{AB}^{(\p)}$. 
The condition $K_{\p}(V,g) \ge 0$ provides a sufficient condition to guarantee secure communication. Nevertheless, our results are in good agreement with other exact numerical approaches \cite{Ghalaii}, proving the bound (\ref{eq:K NotOpt}) to be tight, especially in the long-distance regime $\kappa d \gg 1$.

Thus, in our approach it suffices to compute the CM $\Gamma_{AB}^{(\p)}$ associated with $\rho_{AB}^{(\p)}$ to perform the security analysis. Straightforward calculations lead to (see Appendix~\ref{app: NLA})
\begin{align}\label{eq: GammaAB|p}
    \Gamma_{AB}^{(\p)} = \begin{pmatrix} V_{\p}(V,g)\, \Id_2 & Z_{\p}(V,g) \, \sigmaz \\[1ex] Z_{\p}(V,g) \, \sigmaz & W_{\p}(V,g) \, \Id_2 \end{pmatrix} \, .
\end{align}
The expressions of $P_{\p}(V,g)$, $V_{\p}(V,g)$, $W_{\p}(V,g)$ and $Z_{\p}(V,g)$ are clumsy and thus only reported in Appendix~\ref{app: NLA}. We compute the mutual information and the Holevo information following the procedure described in Sec.~\ref{sec: GG02} by substituting $\Gamma_{AB} \rightarrow \Gamma_{AB}^{(\p)}$ and optimize Eq.~(\ref{eq:K NotOpt}) over the free parameters, obtaining the KGRs
\begin{align}\label{eq: K Opt}
K_{\rm p}(g)= \max_{V} \, K_{\rm p}(V,g)  \, , \quad (\p=\a,\b) \,,
\end{align}
for a fixed $g$, together with the corresponding modulation $V_{\rm opt}^{(\p)}(g)$, and
\begin{align}\label{eq: K Opt}
K_{\rm p}= \max_{V,g} \, K_{\rm p}(V,g)  \, , \quad (\p=\a,\b) \,,
\end{align}
if $g$ can be optimized too, with the associated optimized parameters $V_{\rm opt}^{(\p)}$ and $g_{\rm opt}^{(\p)}$.
\par
We note that in the SPC scheme there always exists a local maximum for $\tau=1$, in which case the SPC performs as the identity operator, allowing to retrieve the results of the original protocol. However, for a more fair comparison with the QS, in the optimization procedure we have neglected this point and restricted maximization over the interval $0\le\tau \le 1/2$ for which the corresponding gain is $g\geq 0$, as shown in Appendix~\ref{app: NLA}.

\section{Security analysis}\label{sec: Security}
\begin{figure}[t]
\includegraphics[width=0.9\columnwidth]{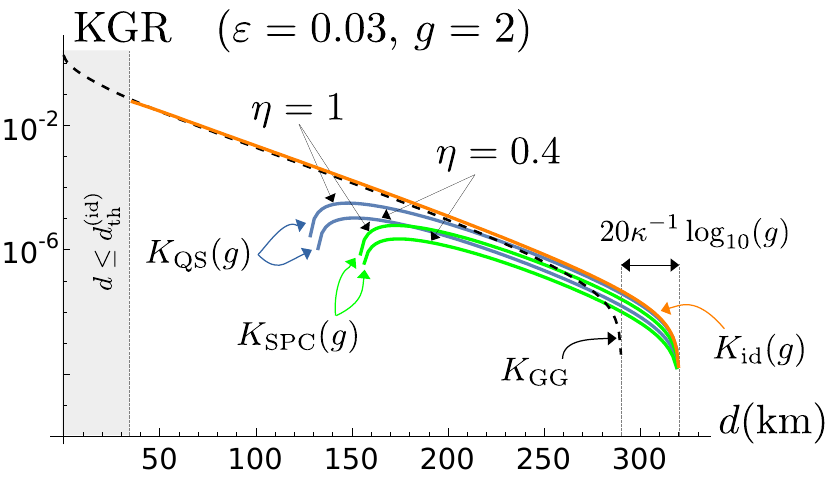}\\[2ex]
\includegraphics[width=0.9\columnwidth]{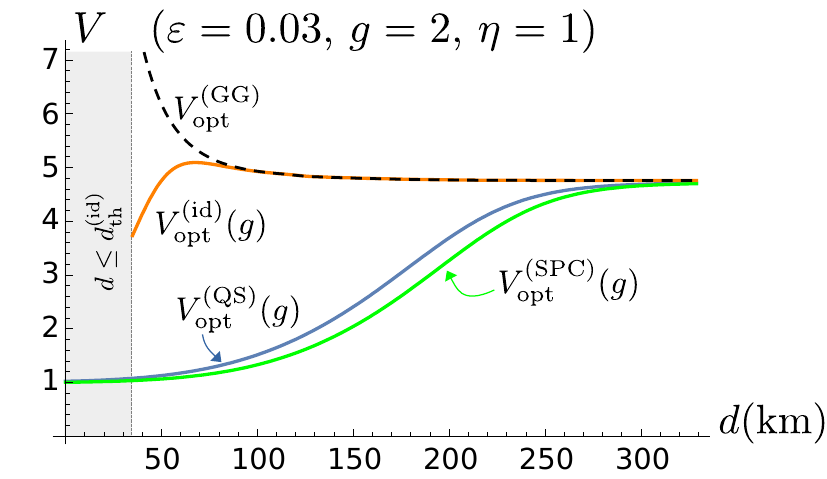}
\caption{
(Top) Log plot of the KGRs $K_{\p}(g)$ for different values of the quantum efficiency $\eta$ and $K_{\id}(g)$ as functions of the distance $d$ in ${\rm km}$. The dashed line is the KGR of the original protocol. (Bottom) Plot of the optimized (input) modulations $V_{\rm opt}^{(\p)}(g)$ and $V_{\rm opt}^{(\id)}(g)$ as a function of the distance $d$ in ${\rm km}$ for $\epsilon=0.03$. In both the plots, the shaded region represents the regime $d\le d_{\rm th}^{(\id)}$, where ideal NLAs generate an unphysical un-normalizable state (see the text for details). We set $\beta=0.95$, $\epsilon=0.03$ and $g=2$.
}\label{fig:03-KGR-Nopt}
\end{figure}
In this section we compare the KGRs of all the schemes under investigation, for the two cases of fixed or optimized gain.
 
\subsection{KGR with fixed gain $g$}\label{sec: gFixed}
For a fixed $g$, the optimized KGRs are depicted in Fig.~\ref{fig:03-KGR-Nopt} (top panel) for $\epsilon>0$. 
As emerges from the plot, NLAs are fundamental in the long-distance regime, as for large $d$ all the NLA-assisted protocols beat the KGR~(\ref{eq: KGR GG02}) of the original protocol. 
The ideal NLA increases the maximum transmission distance by the amount $(20 \log_{10} g)/\kappa$, since for $T\ll 1$ the effective transmissivity in Eq.~(\ref{eq: IdealPar}) is $T_\id \approx g^2 T$ \cite{Blandino}. 
Remarkably, also the physical NLA-assisted protocols achieve the same maximum transmission distance.
Moreover, the presence of inefficient conditional detection reduces the value of the KGRs, still maintaining the same increase in distance even 
for the realistic values of practical CV-QKD systems where $0.4\le \eta\le 0.6$ \cite{Lodewyck_2005, Lodewyck_2007}.
\par
In fact, by expanding the CM~(\ref{eq: GammaAB|p}) in the long-distance regime where $T\ll 1$ up to the first order in $T$, we have:
\begin{subequations}\label{eq: apprCM-Nopt}
\begin{align}
V_{\p}(V,g) &=  V + O(T) \, ,  \\[1.ex]
W_{\p} (V,g) &=  g^2 T (V + \chi) + O(T^2)\, , \\[1.ex]
Z_{\p}(V,g) &=  \sqrt{g^2 T} \, Z + O(T^{3/2}) \, , \quad (\p=\a,\b) \, ,
\end{align} 
\end{subequations}
corresponding to the CM of a GG02 scheme with transmissivity $g^2 T$, consistently with the ideal case.
The success probabilities read 
\begin{equation}
P_{\p}(V,g) \approx P_{\p}(g)=\eta \tau_{\p}(g) \, ,
\end{equation}
and, being $P_{\b}(g) \le P_{\a} (g)$, we have $K_{\b}(g) \le K_{\a} (g)$. In turn, a quantum efficiency $\eta\le 1$ only reduces the success probability and rescales the KGR, without preventing long-distance secure communication.
For completeness, we report the (input) optimized modulations in the bottom panel of Fig.~\ref{fig:03-KGR-Nopt}. Despite the different behaviour at small distances, for large $d$ all the protocols converge to the same asymptotic value, not depending on $\epsilon$. Numerical calculations have also shown that $V_{\rm opt}^{(\p)}(g)$ does not depend on the quantum efficiency.

\begin{figure}[t]
\includegraphics[width=0.9\columnwidth]{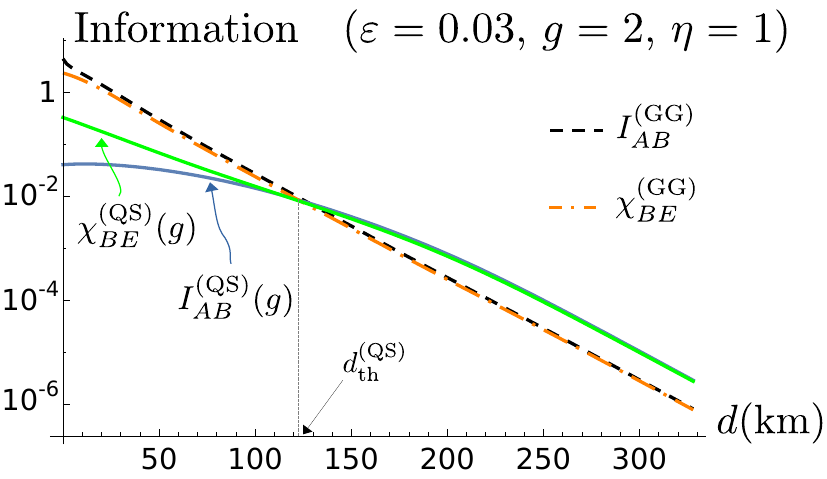}\\[2.ex]
\includegraphics[width=0.9\columnwidth]{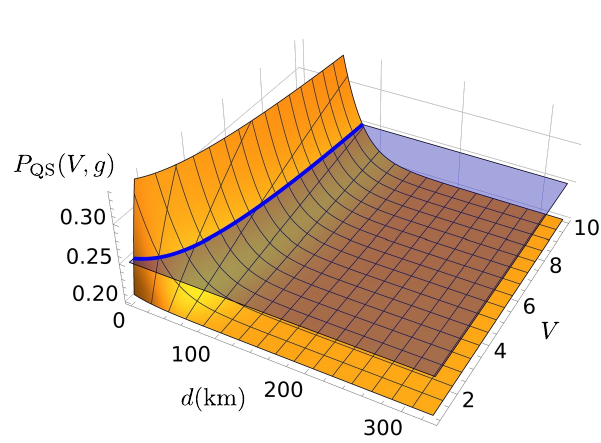}
\caption{
(Top) Log plot of $I_{AB}^{(\a)}(g)$ and $\chi_{BE}^{(\a)}(g)$ (solid lines) and $I_{AB}^{(\rm GG)}$ (dashed line) and $\chi_{BE}^{\rm (GG)}$ (dash-dotted line) as a function of the distance $d$ in ${\rm km}$. 
(Bottom) Plot of the success probability $P_\a(V,g)$ as a function of the distance $d$ and the modulation variance $V$. The horizontal plane refers to the value $1/g^2$: when $P_\a(V,g) > 1/g^2$, the $\a$ do not perform noiseless amplification. In both the pictures we set $\beta=0.95$, $\epsilon=0.03$, $g=2$ and $\eta=1$.
}\label{fig:04-ShortD}
\end{figure}

We note that in the short-distance regime, where $T \approx 1$ or, equivalently, $\kappa d \ll 1$, both the physical NLAs are useless, since we obtain negative KGR up to a threshold distance $d_{\rm th}^{(\p)}$, $\p=\a,\b$. In this regime, the CM~(\ref{eq: GammaAB|p}) cannot be recasted in the form of Eq.~(\ref{eq: Gamma_GG02}), and, as displayed in the top panel of Fig.~\ref{fig:04-ShortD} for the $\a$ case, both the mutual information $I_{AB}^{(\p)} (g)= I_{AB}^{(\p)} (V_{\rm opt}^{(\p)}(g),g)$ and the Holevo information $\chi_{BE}^{(\p)} (g)= \chi_{BE}^{(\p)} (V_{\rm opt}^{(\p)}(g),g)$ are lower than their GG02 counterparts $I_{AB}^{\rm (GG)}$ and $\chi_{BE}^{\rm (GG)}$, respectively. Moreover, for $\epsilon>0$ we have $I_{AB}^{(\p)}(g) \le \chi_{BE}^{(\p)}(g)$, leading to a negative KGR which inhibits secure communication.
This effect may be understood by considering the success probability $P_\p(V,g)$ of the proposed physical NLAs, plotted in the bottom panel of Fig.~\ref{fig:04-ShortD} for the $\a$ case.  Analogous considerations hold for $\b$.
When $P_\p(V,g)>1/g^2$ the $\p$ scheme does not implement a true NLA \cite{Blandino,Ghalaii}, and the amplification process introduces a unavoidable noise on the quadrature variances, becoming a further resource for Eve's attack.  
Accordingly, for $\kappa d \ll 1$ the optimization procedure leads to low modulation variances  $V_{\rm opt}^{(\p)}(g) \approx 1$, resulting in a lower mutual information with respect to the GG02 scheme and in a negative KGR.
On the other hand, for $\kappa d \gg 1$, $V_{\rm opt}^{(\p)} \approx V_{\rm opt}^{\rm (GG)}$ and both $I_{AB}^{(\p)}$ and $\chi_{BE}^{(\p)}$ outperform the GG02 protocol. In turn, between the short- and long-distance regimes, we identify the threshold distance such that $K_\p(g)\ge0$ for $d\le d_{\rm th}^{(\p)}$.

\begin{figure}[t]
\includegraphics[width=0.9\columnwidth]{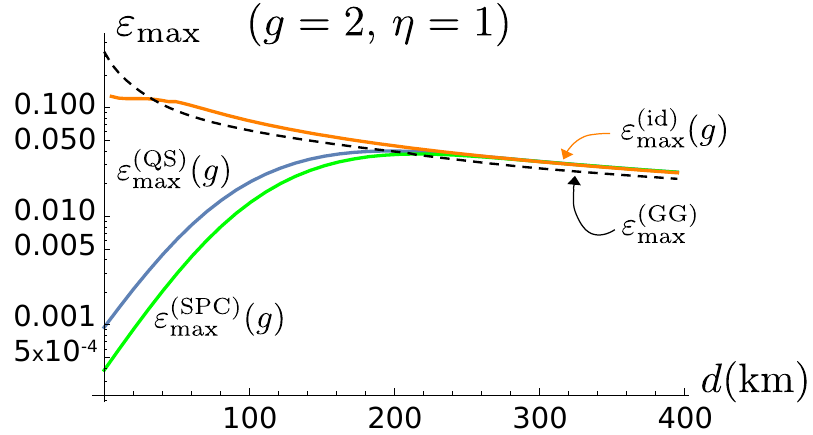}
\caption{
Log plot of the maximum tolerable excess noise $\epsilon_{\rm max}^{(\id)}(g)$ and $\epsilon_{\rm max}^{(\p)}(g)$, $\p=\a,\b$, as a function of the distance $d$ in ${\rm km}$. The black dashed line corresponds to the $\epsilon_{\rm max}$ of the original protocol. We set $\beta=0.95$ and $\eta=1$.}\label{fig:05- epsilonNOpt}
\end{figure}

Finally, in Fig.~\ref{fig:05- epsilonNOpt} we plot the maximum tolerable excess noise (MTEN) $\epsilon_{\rm max}$ as a function of the distance $d$: it represents the maximum value of $\epsilon$ still leading to a positive KGR.
For the original protocol, $\epsilon_{\rm max}$ is a decreasing function of $d$. The behaviour is rather different for the NLA-assisted protocols. In the presence of ideal NLA the MTEN $\epsilon_{\rm max}^{(\id)}(g)$ for $d \lesssim 40$~km is lower than the original protocol due to the limitation imposed by ~(\ref{eq: Condition}). However, for larger distances we have $\epsilon_{\rm max}^{(\id)}(g) > \epsilon_{\rm max}$.
On the contrary, the MTEN associated with the physical NLAs, namely $\epsilon_{\rm max}^{(\p)}(g)$, is not a monotonous function of $d$: it is an increasing function of $d$ approaching $\epsilon_{\rm max}^{(\id)}$. A quantum efficiency $\eta\le1$ does not affect the value of $\epsilon_{\rm max}^{(\p)}$, consistently with the previous discussions.
As a consequence, for fixed $g$, in the long-distance regime the physical NLAs guarantee the same performance of the ideal NLA.

\subsection{KGR with optimized gain $g$}
\begin{figure}[t]
\includegraphics[width=0.9\columnwidth]{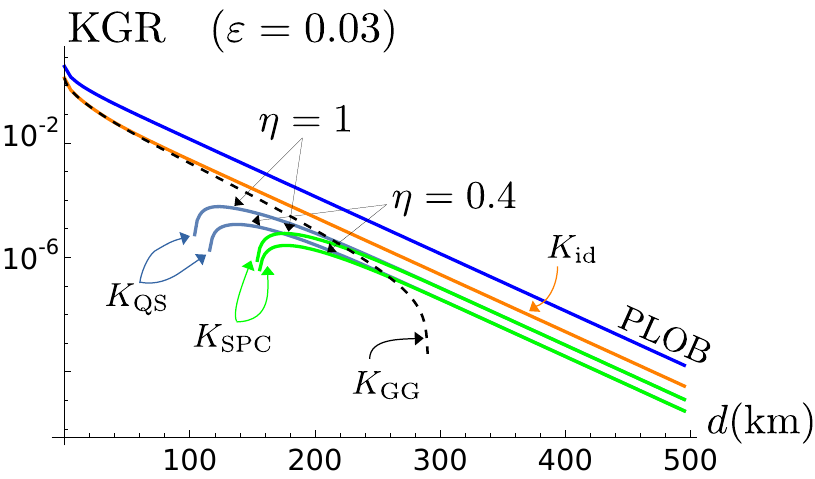}\\[2ex]
\includegraphics[width=0.9\columnwidth]{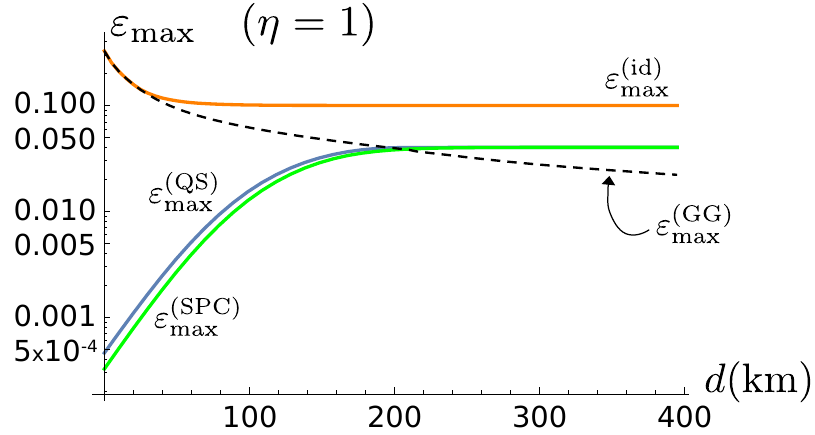}
\caption{
(Top) Log plot of the KGRs $K_{\p}$, $\p=\a,\b$, and $K_{\id}$ as a function of the distance $d$ in ${\rm km}$, for different values of the quantum efficiency $\eta$ and $\epsilon=0.03$ and with optimized gain $g$. The dashed line is the KGR of the original protocol and the upper line is the PLOB bound~(\ref{eq:PLOB}). (Bottom) Log plot of the maximum tolerable excess noises $\epsilon_{\rm max}^{(\id)}$ and $\epsilon_{\rm max}^{(\p)}$, $\p=\a,\b$, as a function of the distance $d$ in ${\rm km}$, for $\eta=1$. The black dashed line corresponds to the $\epsilon_{\rm max}$ of the original protocol. In both the pictures we set $\beta=0.95$.
}\label{fig:06-KGR-Opt}
\end{figure}
The situation is rather different if we can also optimize the gain $g$ associated with the NLAs, as reported in Fig.~\ref{fig:06-KGR-Opt} (top panel). 
Firstly, in the short-distance regime the physical NLAs still exhibits a threshold distance to obtain a positive KGR, differently from the ideal amplifier. 
Secondly, all the NLA-assisted protocols allow to reach arbitrary large distances, but the ideal amplifier outperforms the physical ones. As before, a quantum efficiency still rescales the KGR. However, differently from Sec.~\ref{sec: gFixed}, in the long distance regime $\kappa d \gg 1$, $K_{\a}$ and $K_{\b}$ are almost identical, proving SPC as a feasible alternative to QS. 
\par
We also remark that in the long-distance regime both $K_{\id}$ and $K_{\p}$, $\p=\a,\b$, are proportional to the Pirandola–Laurenza–Ottaviani–Banchi (PLOB) bound \cite{PLOB}
\begin{align}\label{eq:PLOB}
K_{\rm max} = - \log_2 \big[ (1-T) T^{n_\epsilon} \big] - G(n_\epsilon) \, ,
\end{align}
which represents the maximum KGR achievable with the considered repeaterless thermal-loss channel, thus resulting in nearly optimal strategies. 
\par
\begin{figure}[t]
\includegraphics[width=0.9\columnwidth]{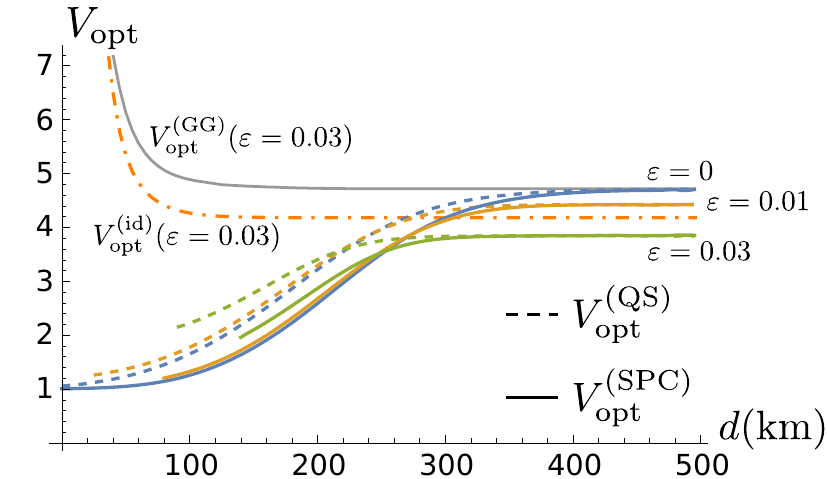}\\[2ex]
\includegraphics[width=0.9\columnwidth]{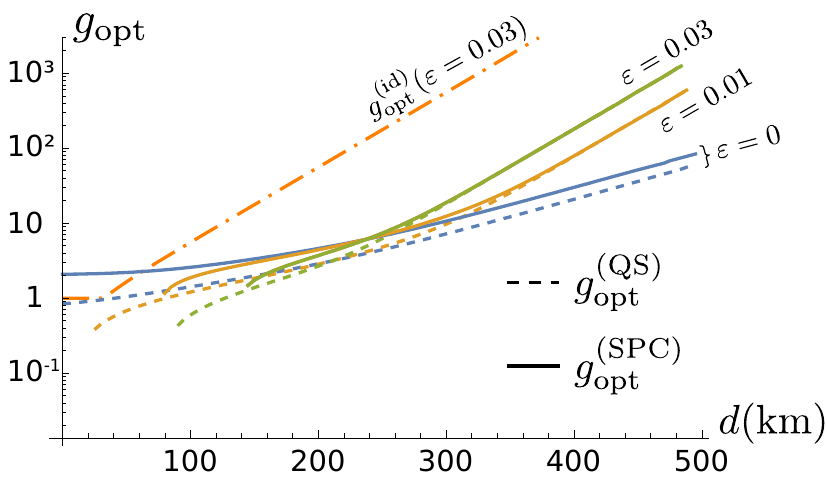}
\caption{(Top) Plot of $V_{\rm opt}^{(\p)}$, $\p=\a,\b$, as a function of the distance $d$ in ${\rm km}$, for different values of excess noise $\epsilon$. The upper gray and the dash-dotted lines represent the optimized modulation for the original and the ideal NLA-assisted protocols, respectively, for $\epsilon=0.03$. (Bottom) Log plot of $g_{\rm opt}^{(\p)}$, $\p=\a,\b$, as a function of the distance $d$ in ${\rm km}$, for different values of excess noise $\epsilon$. The plots have been performed only for the distances such that $K_{\p}>0$, $\p=\a,\b$. We set $\beta=0.95$ and $\eta=1$.}\label{fig:07-OptValues}
\end{figure}

Furthermore, in Fig.~\ref{fig:07-OptValues} we report the optimized parameters $V_{\rm opt}^{(\p)}$ and $g_{\rm opt}^{(\p)}$. 
The modulation $V_{\rm opt}^{(\p)}$ has a different behavior with respect to Sec.~\ref{sec: gFixed}, being an $\epsilon$-dependent growing function of $d$. On the contrary, the modulations of the original and the ideal NLA-assisted protocols are decreasing functions of $d$ converging to an asymptotic value not depending on $\epsilon$, as for the case of fixed $g$.
Instead, the optimized gains $g_{\rm opt}^{(\id)}$ and $g_{\rm opt}^{(\p)}$ grow exponentially with $d$ in the long-distance regime. However, if $\epsilon=0$ this exponential scaling is not reached yet for the physical NLAs within the considered range of distances $d\le 500 \, \rm{km}$.

Finally, in the bottom panel of Fig.~\ref{fig:06-KGR-Opt} we plot the MTENs as a function of $d$.
Differently from Sec.~\ref{sec: gFixed}, the MTEN associated with the physical NLAs, namely $\epsilon_{\rm max}^{(\p)}$, do not achieve the performance of the ideal one, $\epsilon_{\rm max}^{(\id)}$. Actually, both these MTENs outperform the original protocol and saturate to a value  $\epsilon_\infty$ as $\kappa d \gg1$. However, the saturation value of the physical NLAs, namely $\epsilon_\infty^{(\p)}\approx 0.04$, is lower than the ideal NLA one, that is $\epsilon_\infty^{(\id)} \approx 0.1$ (see Fig.~\ref{fig:06-KGR-Opt}). The numerical results also show that a quantum efficiency $\eta\le1$ does not affect the value of $\epsilon_\infty^{(\p)}$, consistently with the previous findings. 
\par
\begin{figure}[t]
\includegraphics[width=0.9\columnwidth]{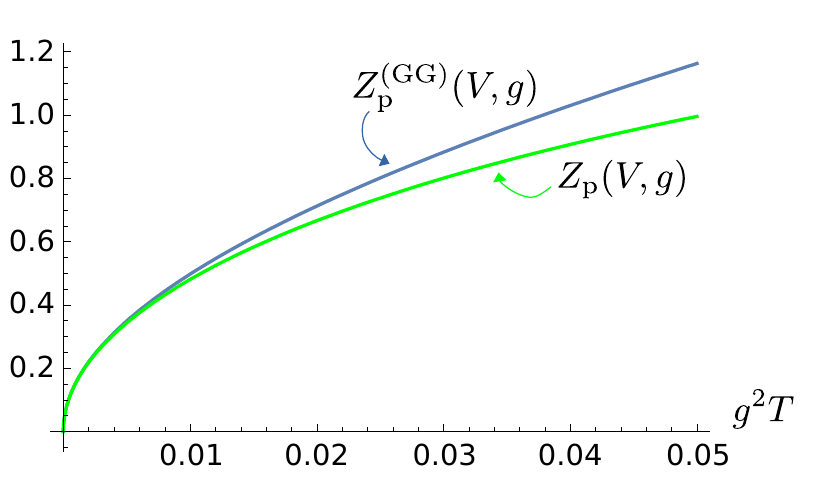}\\[2ex]
\includegraphics[width=0.9\columnwidth]{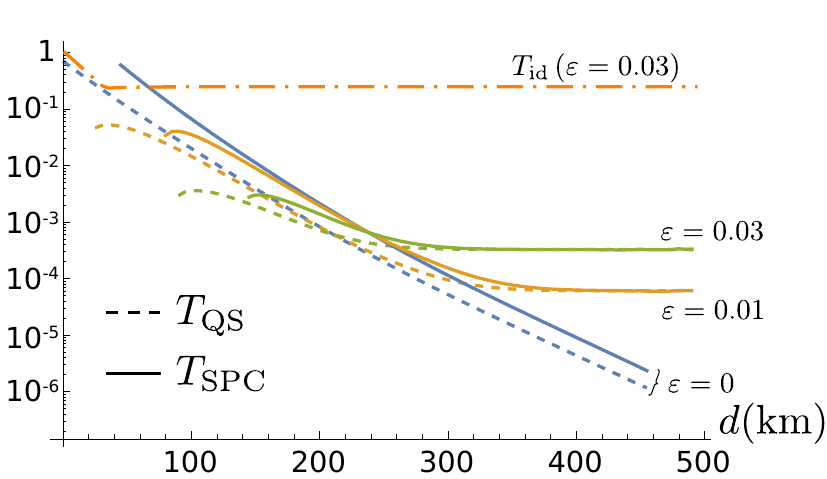}
\caption{(Top) Plot of $Z_{\p}(V,g)$ and $Z_{\p}^{\rm (GG)}(V,g)$, $\p=\a,\b$, as a function of $g^2 T$ for $\epsilon=0.03$ and $V=4$. (Bottom) Log plot of the effective transmissivity $T_{\p}$, $\p=\a,\b$, as a function of the distance $d$ in ${\rm km}$, for different values of excess noise $\epsilon$.  The plot have been performed only for the distances such that $K_{\p}>0$. In both the pictures we set $\beta=0.95$ and $\eta=1$.}\label{fig:08-Teff}
\end{figure}

The difference between ideal and physical NLAs emerges by expanding the CM~(\ref{eq: GammaAB|p}) in the long-distance regime $T\ll 1$ up to the first order, keeping all the contributions of $O(g^2 T)$, due to the fact that $g_{\rm opt}^{(\p)} \gg 1$, and neglecting the other terms:
\begin{subequations}\label{eq: apprCM}
\begin{align}
V_{\p}(V,g) & \approx V + \delta V_{\p} \, ,  \\[1.ex]
W_{\p} (V,g)& \approx T_{\p} \big[V_{\p} (V,g) + \chi_{\p} \big] \, ,  \\[1.ex]
Z_{\p}(V,g) & \approx \frac{T_{\p}}{\sqrt{g^2 T}} \, Z \, , \qquad (\p=\a,\b) \, ,
\end{align}
\end{subequations}
where $\delta V_{\p} =T_{\p} Z^2/2$. $T_{\p}$ represents the effective transmissivity
\begin{align}
T_{\p}= \frac{g^2 \, T}{1+g^2 T (V+\epsilon-1)/2} \, , 
\end{align} 
while $\chi_{\p}= (1- T_{\p})/T_{\p} + \epsilon_{\p}$, with the effective excess noise 
\begin{align}
\epsilon_{\p}= \epsilon- \delta V_{\p} \, .
\end{align}
Employing a physical NLA is then equivalent to considering an effective channel of higher transmissivity $T_{\p}\ge T$ and lower excess noise $\epsilon_{\p}\le \epsilon$.
Nevertheless, the correspondence with a GG02 protocol does not occur anymore, as the correlation term $Z_{\p}(V,g)$ does not coincide with the one expected for a GG02 scheme, namely,
\begin{align}
Z_{\p}^{\rm (GG)}(V,g) = \sqrt{T_{\p} \, \big[V_{\p}(V,g)^2-1 \big]} \, ,
\end{align}
but rather
\begin{align}
Z_{\p}(V,g)\le Z_{\p}^{\rm (GG)}(V,g) \, ,
\end{align}
as depicted in Fig.~\ref{fig:08-Teff} (top panel). We have $Z_{\p}(V,g)\approx Z_{\p}^{\rm (GG)}(V,g)$ only if $g^2 T \ll 1$. As a consequence, the analogy with the ideal-NLA assisted protocol in Eq.~(\ref{eq: IdealPar}) is broken.

Now, the optimization procedure described above leads to exponential gains $g_{\rm opt}^{(\id)}$ and $g_{\rm opt}^{(\p)}$ for the ideal and physical NLAs, respectively, such that the product $g^2 T$ is kept constant for $\kappa d \gg 1$. Consequently, the effective transmissivities $T_{\id}$ and $T_{\p}$ saturate, as shown in the bottom panel Fig.~\ref{fig:08-Teff}.
In turn, also the mutual information and the Holevo information saturate and the corresponding KGRs~(\ref{eq: KGR ideal Opt2}) and~(\ref{eq: K Opt}) turn out to be proportional only to the success probability of the NLAs, namely:
\begin{align}
K_{\id} \propto \frac{1}{\big(g_{\rm opt}^{(\id)}\big)^2} \propto T\, ,
\end{align}
and 
\begin{align}\label{eq: K asymp}
K_{\p} \propto P_{\p}
\approx \frac{\eta T }{2 T_{\p}} \bigg[ 1+ T_{\p}(V_{\p}+ \chi_{\p})\bigg]\, ,
\end{align}
with $P_{\p}=P_{\p}\big(V_{\rm opt}^{(\p)}, g_{\rm opt}^{(\p)}\big)$ and $V_{\p}= V_{\p}\big(V_{\rm opt}^{(\p)}, g_{\rm opt}^{(\p)}\big) $, decreasing linearly with $T$ and thus guaranteeing $K_{\p}>0$ for $\kappa d \gg 1$. The same linear scaling is achieved by the PLOB bound if $T \ll 1$:
\begin{align}
K_{\rm max} \approx T \frac{2-\epsilon[1- \ln(\epsilon/2)]}{2 \ln 2} \, , 
\end{align} 
which proves both all the NLA-assisted protocols to be nearly optimal.
Furthermore, as in Sec.~\ref{sec: gFixed} a quantum efficiency $\eta\le 1$ only rescales the KGR and does not introduce any maximum transmission distance.
\par
Moreover, the saturation value of $T_{\p}$ determines the difference between ideal and physical NLAs. Indeed, if $\epsilon_{\p}$ is small we have $T_{\p}\ll 1$ and the physical NLA-assisted protocols approximate a GG02 protocol with the effective channel parameters $T_{\p}$ and $\epsilon_{\p}$. By increasing the excess noise further, we have $T_{\p}\NOTll 1$ and $Z_{\p}(V,g)\le Z_{\p}^{\rm (GG)}(V,g) $, the state shared between Alice and Bob is less correlated and the protocol deviates more and more from GG02. This implies the reduced asymptotic maximum tolerable excess noise with respect to the ideal case.

\section{Conclusions}\label{sec: Concl}
In this paper we have addressed the exploitation of NLAs to achieve long-distance CV-QKD in the presence of a non-unit reconciliation efficiency and a non-null excess noise of the channel. We have considered both the ideal amplifier and two approximated physical realizations, namely, QS and SPC, in the presence of inefficient conditional on-off detection. We have discussed two alternative scenarios of either fixed or optimized NLA gain and showed that in the former case employing a NLA increases the maximum transmission distance by $(20 \log_{10} g)/\kappa$, whereas in the latter one NLAs allow to reach arbitrary large distances, provided the excess noise of the channel to be sufficiently low. 
Furthermore, we have proved both the physical NLA-assisted protocols to be robust if $\eta \le 1$, showing that the quantum efficiency only rescales the KGR without preventing long-distance communication.
\par
The results obtained offer a further strategy to overcome the practical limitations in CV-QKD and quantifies the degradation of performance produced by inefficient conditional detection. Moreover, they provide new perspectives for the applications of NLAs in realistic conditions for both one-way communication and end-to-end communication over quantum repeater chains \cite{Rep1, Rep2, Rep3}.

\section*{Acknowledgements}
We thank M.~G.~A. Paris for useful comments.
This work has been partially supported by MAECI, Project No.~PGR06314 ``ENYGMA'' and by University of Milan, Project No.~RV-PSR-SOE-2020-SOLIV ``S-O PhoQuLis''.

\appendix

\section{Brief review of the phase-space formalism}\label{app: Gauss}
As discussed in the main text, to perform the analysis of the continuous-variable quantum key distribution (CV-QKD) protocol we exploit the phase-space formalism \cite{Olivares_2012, Ferraro_Olivares_Paris_2005}.
We consider a $n$-mode bosonic system, described by the bosonic operators $a_k$ satisfying the canonical commutation relations $[a_k,a_l]=0$, $[a_k,a_l\dag]=\delta_{kl}$, and by the quadrature operators
\begin{align}
    q_k= a_k+ a_k\dag \quad \mbox{and} \quad p_k= i(a_k\dag-a_k) \, ,
\end{align}
such that $[q_k,p_l]=2 i \delta_{kl}$. All quantities are expressed in shot noise units. A more compact notation is obtained by introducing the vector operators ${\bf a}= (a_1,a_2, \ldots, a_n)^\mathsf{T}$ and ${\bf r}= (q_1,p_1,q_2,p_2, \ldots, q_n,p_n)^\mathsf{T}$. 

\subsection{Quantum states}
According to Glauber's formula \cite{Olivares_2012, Ferraro_Olivares_Paris_2005}, any $n$-mode quantum state of radiation $\rho$ writes:
\begin{align}
\rho = \int \frac{d^2 \bmalpha}{\pi^n} \chi(\bmalpha) D_{\bf a}(\bmalpha)\dag \, ,
\end{align}
where $\bmalpha= (\alpha_1,\alpha_2, \ldots, \alpha_n)^\mathsf{T} \in \mathbb{C}^n$ and 
\begin{align}
D_{\bf a}(\bmalpha) = \bigotimes_{k=1}^{n} D_{a_k}(\alpha_k) \, ,
\end{align}
where $D_{a_k}(\alpha_k)$ is the displacement operator acting on mode $a_k$, namely,
\begin{align}
D_{a_k}(\alpha_k)= \exp(\alpha_k a_k\dag - \alpha_k^* a_k) \, .
\end{align}
Some useful properties of the displacement operator are reported below:
\begin{subequations}\label{eq: properties}
\begin{align}
&D_{\bf a}(\bmalpha_1) D_{\bf a}(\bmalpha_2) = D_{\bf a}(\bmalpha_1+\bmalpha_2) \, , \quad \bmalpha_1,\bmalpha_2 \in\mathbb{C}^n \,,  \\
&D_{\xi \bf a}(\bmalpha)= D_{\bf a}(\xi \bmalpha)\, , \quad \xi \in\mathbb{R} \,,  \\
&\Tr\big[D_{\bf a}(\bmalpha)\big] = \pi^n \delta^{(n)} (\bmalpha) \, ,
\end{align}
\end{subequations}
$\delta^{(n)} (\bmalpha)$ being the complex $n$-mode Dirac delta distribution.
\par
Finally, the function
\begin{align}
\chi(\bmalpha) = \Tr\big[\rho D_{\bf a}(\bmalpha)\big]
\end{align}
is the characteristic function associated with $\rho$.
In particular, a quantum state $\rho_G$ exhibiting a Gaussian characteristic function is said to be a \textit{Gaussian state}, namely,
\begin{align}\label{eq: chi_Gauss}
    \chi(\bmalpha)=  \exp \bigg[ 
    - \displaystyle \frac12 \tilde{\bmalpha}^\mathsf{T} \, \boldsymbol\sigma \, \tilde{\bmalpha} - i \tilde{\bmalpha}^\mathsf{T} {\bf X}
    \bigg] \, ,
\end{align}
where $\tilde{\bmalpha}= (\Re \alpha_1, \Im \alpha_1,\Re \alpha_2, \Im \alpha_2, \ldots, \Re \alpha_n, \Im \alpha_n) \in \mathbb{R}^{2n}$, 
\begin{align}
    {\bf X}= \Tr[\rho_G \, {\bf r}]
\end{align}
is the first moment vector and
\begin{align}
    \boldsymbol\sigma = \frac12 \Tr \bigg[ \rho_G \, \big\{({\bf r}-{\bf X}),({\bf r}-{\bf X})^\mathsf{T}\big\} \bigg] \, 
\end{align}
is the $2n\times 2n$ covariance matrix (CM) where $\{A,B\}=AB+BA$ is the anti-commutator of $A$ and $B$.
Thus, a Gaussian state is completely characterized by its prime moments and its covariance matrix.
\par
Moreover, for any pair of generic operators $O_1$ and $O_2$ acting on the Hilbert space of $n$ modes the \textit{trace rule} holds:
\begin{align}
\Tr[O_1 O_2] = \int \frac{d^2\bmalpha}{\pi^n} \chi_{O_1}(\bmalpha) \chi_{O_2}(-\bmalpha) \, ,
\end{align}
$\chi_{O_{1(2)}}(\alpha)$ being the characteristic function of $O_{1(2)}$, respectively.
As an example, for a single radiation mode $a$, we choose $O_1=D_a(\alpha)$ and $O_2=q^2_a=(a+a\dag)^2$ and obtain \cite{Ghalaii}:
\begin{widetext}
\begin{align}\label{eq: CovDisp}
\Tr\big[D_a(\alpha) q^2_a\big]= e^{-(x^2+y^2)/2} \Bigg[\pi \delta^{(2)} (\alpha)+ 2 \pi y \delta(x) \frac{d}{d y} \delta(y) - \pi \delta(x) \frac{d^2}{d y^2} \delta(y)\Bigg] \, ,
\end{align}
\end{widetext}
where $\alpha=x+i y$ and $\delta(x)$ is the Dirac delta distribution.

\subsection{Conditional measurements}
In the paper we also discuss the case of conditional measurements. We consider a bipartite system $AB$, where subsystems $A$ and $B$ are composed of $n_{A}$ an $n_B$ modes, respectively. In the vector notation we have ${\bf a}= ({\bf a}_A,{\bf a}_B)$. We consider a bipartite quantum state $\rho_{AB}$ with characteristic functions $\chi_{AB}(\bmalpha)= \chi_{AB}(\bmalpha_A, \bmalpha_B)$.
We now perform a quantum measurement on subsystem $B$, described my means of the positive-operator-valued measurement (POVM) $\{\Pi_{{\bf r}_m}\}_{{\bf r}_m}$, whose effects are associated with the characteristic function $\chi_{{\bf r}_m}(\bmalpha_B)$.
By applying the trace rule, the conditional state on $A$ reads:
\begin{align}\label{eq: CondState}
\rho_{A|{\bf r}_m} &= \frac{1}{p({\bf r}_m)} \Tr_B\big[\rho_{AB} \big(\Id_A \otimes \Pi_{{\bf r}_m} \big)\big] \notag \\
&\equiv \frac{1}{p({\bf r}_m)} \int \frac{d^2\bmalpha_A} {\pi^{n_A}}\, \chi_{A|{\bf r}_m}(\bmalpha_A)\,  D_{{\bf a}_A} (\bmalpha_A)\dag\, ,
\end{align}
where:
\begin{align}
\chi_{A|{\bf r}_m}(\bmalpha_A) = 
\int \frac{d^2\bmalpha_B} {\pi^{n_B}}\,  \chi_{AB}(\bmalpha_A,\bmalpha_B)\, \chi_{{\bf r}_m}(-\bmalpha_B) \, ,
\end{align}
and $p({\bf r}_m)$ is the detection probability:
\begin{align}\label{eq: SuccP}
p({\bf r}_m) &= \Tr_{AB} \big[\rho_{AB} \big(\Id_A \otimes \Pi_{{\bf r}_m} \big)\big] \notag \\
&= \Tr_{A} \Bigg[\int \frac{d^2\bmalpha_A} {\pi^{n_A}}\, \chi_{A|{\bf r}_m}(\bmalpha_A)\,  D_{{\bf a}_A} \Bigg]  
= \chi_{A|{\bf r}_m}({\bf 0}) \, .
\end{align}
\par
An interesting results is obtained for Gaussian states and Gaussian measurements. We now assume $\rho_{AB}$ to be a Gaussian state with prime moments $\mathbf{X}=(\mathbf{X}_A,\mathbf{X}_B)$ and CM (written in block form)
\begin{equation}
\boldsymbol\sigma = \begin{pmatrix} \boldsymbol\sigma_A & \boldsymbol\sigma_{AB} \\[1ex]\boldsymbol\sigma_{AB}^\mathsf{T} & \boldsymbol\sigma_B \end{pmatrix} \, .
\end{equation}
Moreover, we consider a Gaussian POVM $\{\Pi_{{\bf r}_m}\}_{{\bf r}_m}$, that is a POVM whose effects have a Gaussian characteristic function with prime moments ${\bf r}_m$ and CM $\boldsymbol\sigma_m$.
Then, the conditional state $\rho_{A|\mathbf{r}_m}$ is still a Gaussian state with CM $\boldsymbol\sigma_{A|\mathbf{r}_m}$ and first moment vector $\mathbf{X}_{A|\mathbf{r}_m}$ given by \cite{Olivares_2012,Ferraro_Olivares_Paris_2005}:
\begin{equation}
    \boldsymbol\sigma_{A|\mathbf{r}_m} = \boldsymbol\sigma_A - \boldsymbol\sigma_{AB} (\boldsymbol\sigma_B+\boldsymbol\sigma_m)^{-1} \boldsymbol\sigma_{AB}^\mathsf{T}\,,
\end{equation}
and
\begin{equation}
    \mathbf{X}_{A|\mathbf{r}_m} = \mathbf{X}_A + \boldsymbol\sigma_{AB} (\boldsymbol\sigma_B+\boldsymbol\sigma_m)^{-1}(\mathbf{r}_m-\mathbf{X}_B)\,,
\end{equation}
respectively.

\section{Security proof of the GG02 protocol}\label{app: EBGG02}
To perform the security analysis of the GG02 protocol in a reverse reconciliation scheme, we shall compute the
KGR: 
\begin{align}
    K= \beta I_{AB}- \chi_{BE} \, , 
\end{align}
$\beta$ being the reconciliation efficiency. 

The mutual information $I_{AB}$ gets the final expression reported in Eq.~(\ref{eq: IAB GG02}), as the Shannon entropy of a multivariate $n$-dimensional Gaussian distribution $\mathcal{N}(\mathbf{\mu},\boldsymbol\sigma)$ with prime moments $\mathbf{\mu}$ and CM $\boldsymbol\sigma$: 
\begin{align}
    \mathcal{G}(\mathbf{x})= \frac{\exp\bigg[-\displaystyle \frac12(\mathbf{x}-\mathbf{\mu})^T \boldsymbol\sigma^{-1} (\mathbf{x}-\mathbf{\mu})\bigg]}{(2\pi)^{n/2} \sqrt{\det(\boldsymbol\sigma)}}
\end{align}
is equal to
\begin{align}
    H[\mathcal{G}] &= -\int d \mathbf{x}~\mathcal{G}(\mathbf{x}) \log_2 \left[ \mathcal{G}(\mathbf{x}) \right]\notag \\
    &=  \frac{1}{2} \big\{
    n \log_2 (2\pi e) + \log_2 \left[\det(\boldsymbol\sigma) \right]
    \big\} \, .
\end{align}
\par
The amount of information extracted by Eve is given by the Holevo information
\begin{align}
    \chi_{BE}= S_E- S_{E|B} \, ,
\end{align}
that can be evaluated as follows.
We assume Eve to purify the system $AB$ shared between Alice and Bob, that is we assume her to collect the fraction of the signal lost due to both the presence of the excess noise and the propagation into the channel such that the global quantum state $\rho_{ABE}$ shared by Alice, Bob and Eve is pure \cite{GarciaPatronCerf_2009,Lodewyck_2007}. As a consequence, we have 
\begin{align}
    S_E= S_{AB}= G\bigg(\frac{d_1-1}{2}\bigg)+G\bigg(\frac{d_2-1}{2}\bigg) \, ,
\end{align}
where $G(x)= (x+1) \log_2 (x+1) - x \log_2 x $
and $d_{1(2)}$ are the symplectic eigenvalues of $\Gamma_{AB}$ \cite{Ferraro_Olivares_Paris_2005, Olivares_2012}. Furthermore, when Bob gets the outcome $x_B$ from homodyne detection and reveals its value, the system $AE$ shared between Alice and Eve becomes pure, thus 
\begin{align}
S_{E|B}= S_{A|B} = G\bigg(\frac{d_3-1}{2}\bigg) \, ,
\end{align}
where $d_3 = \sqrt{\det(\Gamma_{A|B})} $ and
\begin{align}
    \Gamma_{A|B} = \Gamma_A - \Gamma_Z \bigg[ \Gamma_B + \sigmamB \bigg]^{-1} \Gamma_Z^\mathsf{T} \, ,
\end{align}
which is independent of the particular outcome obtained.

\section{Employing quantum scissors (QS) and single-photon catalysis (SPC)}\label{app: NLA}
\begin{figure}
\includegraphics[width=0.99\linewidth]{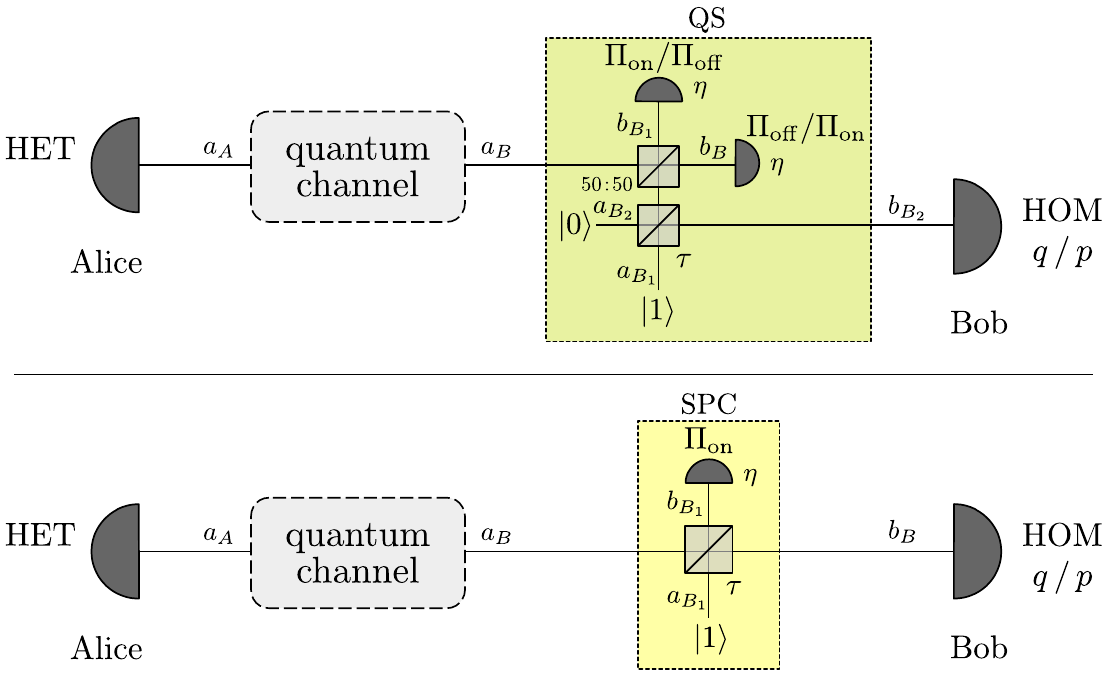}
\caption{Schematic representation of the two physical NLA-assisted protocol discussed in the paper. (Top) Strategy based on quantum scissors (QS);
(bottom) strategy based on single-photon catalysis (SPC).}\label{fig:S01-protocol}
\end{figure}
As discussed in the main text, we perform the security analysis by exploiting the optimality of Gaussian attacks \cite{GaussOpt1, GaussOpt2, GaussOpt3}. If Alice and Bob share a non-Gaussian state $\varrho$, a lower bound of the exact KGR is obtained by considering a Gaussian protocol in which they share the Gaussian state $\varrho_{\rm G}$ with the same CM of $\varrho$. In this section we derive the CM for both the physical noiseless linear amplifiers (NLAs) discussed in the paper, namely the quantum scissors ($\a$) and the single-photon catalysis ($\b$). To do so, we exploit the input-output formalism and the phase-space representation of quantum states.

\subsection{Quantum scissors (QS)}

By following the notation introduced in Fig.~\ref{fig:S01-protocol} (top panel), the protocol employing QS works as follows \cite{Ghalaii}.
Alice prepares the TMSV and injects one mode into the thermal-loss channel, thereafter Bob performs the QS protocol on the received beam. The input modes are ${\bf a} = (\aA, \aB, \aC, \aD)^{\mathsf{T}}$, where $\aA, \aB$ are the modes shared by Alice and Bob after the channel whereas $\aC, \aD$ are the modes exploited locally by Bob for the QS. The global input state reads:
\begin{align}
\rho_{\bf a} = \int \frac{d^2 \bmalpha}{\pi^4} \chi_{\bf a}(\bmalpha) D_{\bf a}(\bmalpha)\dag \, ,
\end{align}
where $\bmalpha=(\alphaA, \alphaB, \alphaC, \alphaD)^{\mathsf{T}}$ and
\begin{align}
\chi_{\bf a}(\bmalpha) = \chi_{\rm G}(\alphaA, \alphaB) \times\big(1-  |\alphaC|^2\big) \, e^{- (|\alphaC|^2 + |\alphaD|^2)/2} \, ,
\end{align}
$\chi_{\rm G}(\alphaA, \alphaB)$ being the Gaussian characteristic function in Eq.~(\ref{eq: chi_Gauss}) with null prime moments and the CM~(\ref{eq: Gamma_GG02}).
\par
The output modes after the mode mixing operations performed by Bob are ${\bf b} = (\bA, \bB, \bC, \bD)^{\mathsf{T}}= {\cal M}_{\a} {\bf a}$, where
\begin{align}
{\cal M}_{\a}= 
\begin{pmatrix}
1 & 0 &  0 & 0 \\
0 & \frac{1}{\sqrt{2}} &  \sqrt{\tau/2} & -\sqrt{(1-\tau)/2} \\[1ex]
0 & -\frac{1}{\sqrt{2}} &  \sqrt{\tau/2} & -\sqrt{(1-\tau)/2} \\[1ex]
0 & 0 &  \sqrt{1-\tau} & \sqrt{\tau}
\end{pmatrix} \, ,
\end{align}
with $\tau= \tau_{\a}(g)= (1+g^2)^{-1}$.
The output state then writes:
\begin{align}
\rho_{\bf b} = \int \frac{d^2 \bmbeta}{\pi^4} \chi_{\bf b}(\bmbeta) D_{\bf b}(\bmbeta)\dag \, ,
\end{align}
where, exploiting the properties in Eq.~(\ref{eq: properties}), $\chi_{\bf b}(\bmbeta)= \chi_{\bf a}({\cal M}_{\a}^{\mathsf{T}} \bmalpha)$.
\par
Finally, Bob performs on-off detection on modes $\bB, \bC$, corresponding to the positive-operator-valued measurement (POVM) $\{\Pi_{\rm off}, \Pi_{\rm on}=\Id-\Pi_{\rm off}\}$, with associated characteristic functions \cite{Ferraro_Olivares_Paris_2005, GaussianOff}
\begin{subequations}
\begin{align}
\chi_{\rm off} (\alpha) &= \frac{1}{\eta} e^{- \frac{2-\eta}{2\eta}|\alpha|^2}\, \\ 
\chi_{\rm on} (\alpha) &= \pi \delta^{(2)}(\alpha)- \chi_{\rm off} (\alpha) \, .
\end{align}
\end{subequations}
The amplification is successful if one of the two detectors gives the outcome ``on" \cite{Ghalaii, NLARalphLund}. In the following we assume to retrieve the couple (on,off), respectively for modes $\bB, \bC$. The post-selected state then equals to:
\begin{align}
\varrho_{\a} = \frac{1}{\widetilde{P}_{\a}} \int \frac{d^2 \betaA}{\pi} \frac{d^2 \betaD}{\pi} \chi_{\a}(\betaA,\betaD) D_{\bA}(\betaA)\dag D_{\bD}(\betaD)\dag  \, ,
\end{align}
where
\begin{align}
\chi_{\a}(\betaA,\betaD)=  \int \frac{d^2 \betaB}{\pi} \frac{d^2 \betaC}{\pi} \chi_{\bf b}(\bmbeta) \chi_{\rm on} (-\betaB) \chi_{\rm off} (-\betaC) \, ,
\end{align}
and
\begin{align}
\widetilde{P}_{\a} &= \Tr \Bigg[ \int \frac{d^2 \betaA}{\pi} \frac{d^2 \betaD}{\pi} \chi_{\a}(\betaA,\betaD) D_{\bA}(\betaA)\dag D_{\bD}(\betaD)\dag \Bigg] \notag \\[1ex]
&= \chi_{\a}(0,0) 
= 2 \frac{8 \eta \tau + (w-1)(3+w)(1+\eta \tau)}{(1+w)^2 (3+w)^2}
\end{align}
is the success probability of this conditional operation, with $w= 1+\eta T (V+\epsilon-1)$. The same results hold if Bob gets the pair (off,on), thus the global success probability of the QS-based NLA is 
$P_{\a}= 2\widetilde{P}_{\a}$.
\par
Finally, we compute the CM associated with the state $\varrho_{\a}$. By exploiting Eq.~(\ref{eq: CovDisp}), we have:
\begin{subequations}
\begin{align}
V_{\a}&= \Tr\big[\varrho_{\a} q_{\bA}^2\big] = - 1 - \frac{{\cal V}_{\a}}{\widetilde{P}_{\a} } \, ,\\[1ex]
W_{\a}&= \Tr\big[\varrho_{\a} q_{\bD}^2\big] = - 1 - \frac{{\cal W}_{\a}}{\widetilde{P}_{\a} } \, , \\[1ex]
Z_{\a}&= \Tr\big[\varrho_{\a} q_{\bA} q_{\bD}\big] = -\frac{{\cal Z}_{\a}}{\widetilde{P}_{\a} } \, ,
\end{align}
\end{subequations}
where
\begin{widetext}
\begin{subequations}
\begin{align}
{\cal V}_{\a}&=\Bigg[\frac{d^2}{dy^2} \Big(e^{-y^2/2}\chi_{\a}(i y,0)\Big)\Bigg]_{y=0} \notag \\
&= 2 (V+1) \Bigg[ \frac{(2+\eta T \epsilon)(1-\eta \tau)}{(1+w)^2}- \frac{8(3+w)+2 \eta T \epsilon(3+w-4\eta\tau)+4\eta \tau (w-5)}{(3+w)^3}\Bigg] \, ,\\
{\cal W}_{\a} &= \Bigg[\frac{d^2}{dv^2} \Big(e^{-v^2/2}\chi_{\a}(0,i v)\Big)\Bigg]_{y=0}
= -4 \, \frac{8 \eta \tau + (w-1)(3+w) [2-(1-\eta)\tau]}{(1+w)(3+w)^2}\, ,\\
{\cal Z}_{\a}&= \Bigg[\frac{d^2}{dy dv} \Big(e^{-(y^2-v^2)/2} \chi_{\a}(i y,i v)\Big)\Bigg]_{y=0, v=0} 
= \sqrt{T Z} \, \frac{8 \eta \sqrt{\tau(1-\tau)}}{(3+w)^2}\, .
\end{align}
\end{subequations}
\end{widetext}
Accordingly, the CM writes:
\begin{align}
    \Gamma^{(\a)}_{AB} =
    \begin{pmatrix} 
V_{\a} \, \Id_2 & Z_{\a} \, \sigmaz\\[1ex]
    Z_{\a} \, \sigmaz & W_{\a} \, \Id_2 
\end{pmatrix} \, .
\end{align}

\subsection{Single-photon catalysis (SPC)}
For SPC we follow the analogous procedure of the previous subsection. The input modes depicted in the bottom panel of Fig.~\ref{fig:S01-protocol} are ${\bf a} = (\aA, \aB, \aC)^{\mathsf{T}}$, where $\aA, \aB$ are the modes shared by Alice and Bob after the channel and $\aC$ is Bob's ancillary mode. The global input state reads:
\begin{align}
\rho_{\bf a} = \int \frac{d^2 \bmalpha}{\pi^3} \chi_{\bf a}(\bmalpha) D_{\bf a}(\bmalpha)\dag \, ,
\end{align}
where $\bmalpha=(\alphaA, \alphaB, \alphaC)^{\mathsf{T}}$  and
\begin{align}
\chi_{\bf a}(\bmalpha) = \chi_{\rm G}(\alphaA, \alphaB) \times e^{- |\alphaC|^2/2} \big(1-  |\alphaC|^2\big) \, ,
\end{align}
$\chi_{\rm G}(\alphaA, \alphaB)$ being the Gaussian characteristic function in Eq.~(\ref{eq: chi_Gauss}) with null prime moments and the CM~(\ref{eq: Gamma_GG02}).
\par
The output modes after the mode mixing operation performed by Bob are ${\bf b} = (\bA, \bB, \bC)^{\mathsf{T}}= {\cal M}_{\b} {\bf a}$, where
\begin{align}
{\cal M}_{\b}= 
\begin{pmatrix}
1 & 0 &  0 \\[0.5ex]
0 & \sqrt{\tau} &  \sqrt{1-\tau} \\[0.5ex]
0 &  -\sqrt{1-\tau} & \sqrt{\tau}
\end{pmatrix} \, ,
\end{align}
with $\tau=\tau_{\b}(g)= \left( 4 + g^2 - g \sqrt{8 + g^2}\right)/8$.
The output state then writes:
\begin{align}
\rho_{\bf b} = \int \frac{d^2 \bmbeta}{\pi^3} \chi_{\bf b}(\bmbeta) D_{\bf b}(\bmbeta)\dag \, ,
\end{align}
where
\begin{align}
\chi_{\bf b}(\bmbeta)= \chi_{\bf a}({\cal M}_{\b}^{\mathsf{T}} \bmalpha) \, .
\end{align}
\par
After the conditional on-off detection on mode $\bC$, the post-selected state reads:
\begin{align}
\varrho_{\b} = \frac{1}{P_{\b}} \int \frac{d^2 \betaA}{\pi} \frac{d^2 \betaB}{\pi} \chi_{\b}(\betaA,\betaB) D_{\bA}(\betaA)\dag D_{\bB}(\betaB)\dag  \, ,
\end{align}
where
\begin{align}
\chi_{\b}(\betaA,\betaB)=  \int \frac{d^2 \betaC}{\pi} \chi_{\bf b}(\bmbeta) \chi_{\rm on} (-\betaC) \, ,
\end{align}
and
\begin{align}
P_{\b} &= \Tr \Bigg[ \int \frac{d^2 \betaA}{\pi} \frac{d^2 \betaB}{\pi} \chi_{\b}(\betaA,\betaB) D_{\bA}(\betaA)\dag D_{\bB}(\betaB)\dag \Bigg]\notag \\[1ex]
&= \chi_{\b}(0,0) = 1-\frac{4(1-\eta \tau) + 2(w-1)(1- \tau)}{[2+(w-1)(1-\tau)]^2}
\end{align}
is the success probability of the SPC, and we introduced the quantity $w= 1+\eta T (V+\epsilon-1)$.
\par
The CM associated with the state $\varrho_{\b}$ reads:
\begin{align}
\Gamma^{(\b)}_{AB} =
\begin{pmatrix} 
V_{\b} \, \Id_2 & Z_{\b} \, \sigmaz\\[1ex]
    Z_{\b} \, \sigmaz & W_{\b} \, \Id_2 
\end{pmatrix} \, .
\end{align}
As for QS, we have:
\begin{align}
V_{\b}&= \Tr\big[\varrho_{\a} q_{\bA}^2\big] = - 1 - \frac{{\cal V}_{\b}}{P_{\b} } \, , \\[1ex]
W_{\b}&= \Tr\big[\varrho_{\a} q_{\bB}^2\big] = - 1 - \frac{{\cal W}_{\b}}{P_{\b} } \, , \\[1ex]
Z_{\b}&= \Tr\big[\varrho_{\a} q_{\bA} q_{\bB}\big] = -\frac{{\cal Z}_{\b}}{P_{\b} } \, ,
\end{align}
and
\begin{widetext}
\begin{subequations}
\begin{align}
{\cal V}_{\b}&=\Bigg[\frac{d^2}{dy^2} \Big(e^{-y^2/2}\chi_{\b}(i y,0)\Big)\Bigg]_{y=0} \notag \\
&= - (V+1) \Bigg[1-2 \, \frac{4+\eta T \epsilon (1-\tau)(1+q-4\eta\tau)+2(1+\eta\tau)(q-1)-4\eta\tau}{(1+q)^3}\Bigg] \, ,\\
{\cal W}_{\b} &= \Bigg[\frac{d^2}{dv^2} \Big(e^{-v^2/2}\chi_{\b}(0,i v)\Big)\Bigg]_{y=0} \notag \\
&= -4 -\tau(r-3)+ 4\, \frac{(q-1)^2 +(r-1)(q-1)(\eta+\tau) +2\tau (r-1) -2 \eta\tau (q-1)+2(w-1)(4-4\tau-\tau^2)}{(1+q)^3}\, ,\\
{\cal Z}_{\b}&= \Bigg[\frac{d^2}{dy dv} \Big(e^{-(y^2-v^2)/2} \chi_{\b}(i y,i v)\Big)\Bigg]_{y=0, v=0} 
= \sqrt{\tau T Z} \Bigg[1-4\, \frac{2+(1+\eta)(q-1)+2\eta(1-2\tau)}{(1+q)^3} \Bigg] \, ,
\end{align}
\end{subequations}
with $q= 1+ \eta T(1-\tau) (V+\epsilon-1)$ and $r= 1+ T (V+\epsilon-1)$.
\end{widetext}


\end{document}